\input epsf
\epsfverbosetrue
\documentstyle[prb,aps,twocolumn]{revtex}
\begin{document}
\title{A very low temperature STM for the local spectroscopy of mesoscopic 
structures}
\author{N. Moussy, H. Courtois and B. Pannetier}
\address{Centre de Recherches sur les Tr\`es Basses 
Temp\'eratures-C.N.R.S. associated to Universit\'e Joseph Fourier,\\ 25 
Ave. des Martyrs, 38042 Grenoble, France} 
\date{\today}
\maketitle
\begin{abstract}
We present the design and operation of a very-low temperature Scanning 
Tunneling Microscope (STM) working at $60 \, mK$ in a dilution refrigerator. 
The STM features both atomic resolution and micron-sized scanning 
range at low temperature. We achieved an efficient 
thermalization of the sample while maintaining a clean surface for 
STM imaging. Our spectroscopic data show unprecedented energy 
resolution. We present current-voltage characteristics and 
the deduced local density of states of hybrid 
Superconductor-Normal metal systems. This work is the first experimental 
realization of a local spectroscopy of mesoscopic structures at 
very low temperature.
\end{abstract}

\section{Introduction}

Mesoscopic effects in metallic structures are 
quantum interference phenomena based on electron phase 
coherence.\cite{Imry} 
This coherence persists over a length $L_{\varphi}$ which increases 
with decreasing temperature and usually reaches about 
$1 \, \mu m$ in clean metals below $1\, K$. In 
particular, much interest was recently devoted to 
the proximity effect occurring in hybrid mesoscopic structures made 
of a Superconductor (S) in clean contact 
with a Normal metal (N).\cite{RevueJLTP} So far, transport measurements have 
been the primary means of investigation of such mesoscopic structures. A 
promising alternative approach is the use of a very-low temperature 
scanning probe. For instance, a Scanning Tunneling Microscope (STM) 
enables one to access the local density of states with atomic 
spatial resolution. With regards to the proximity effect, 
the local modification of the density of states
has been measured at moderately low temperature ($T > 1\, K$) by 
several groups in various kinds of samples.\cite{TakaSTM,PE,Levi,Vinet}

Here, we present the design and operation of a very-low temperature 
STM operating at $60 \, mK$ in a dilution refrigerator and featuring 
a large ($\simeq 6\, \mu m$) scan range. Operating at such very low 
temperatures allows one to gain access to a wide range of phenomena and reduces 
the thermal smearing of the measured density of states. The combination 
of very-low temperature techniques 
with the constraints of scanning probe microscopy is indeed a 
demanding task. This explains the relatively small number of 
such studies. To our knowledge, the only spectroscopic 
measurement at a temperature below $100 \, mK$ was reported in 
$NbSe_{2}$.\cite{Hess} A very-low temperature Atomic Force Microscope 
was operated at $30\, mK$ in a $9\, T$ magnetic field.\cite{Nunes} 
Images of adsorbed $He$ atoms were obtained at $90\, 
mK$.\cite{Fukuyama} Various STMs operating at $300\, mK$ have 
also been recently reported.\cite{He3} 

\begin{figure}
\epsfxsize=8 cm
\epsfbox{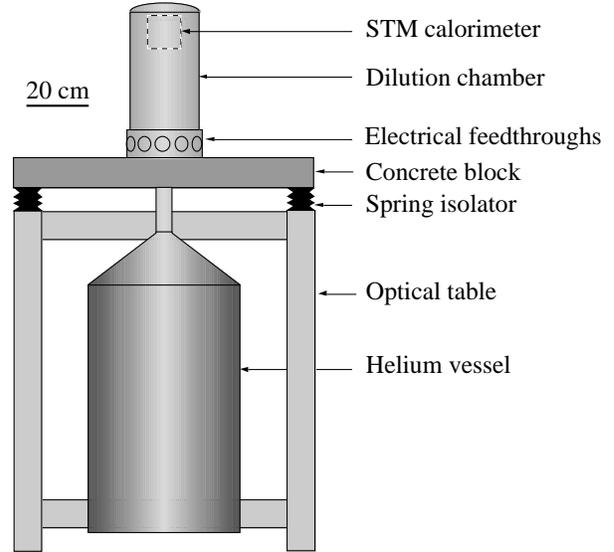}
\caption{Schematics of the cryostat. The cryostat is fixed on a 
concrete block supported by an optical table with 4 spring isolators. The 
pressurized He vessel underneath provides the $^4He$ flow. The 
STM calorimeter is located in the upper part of the dilution cryostat. 
The pumps are located in a nearby acoustically isolated room.}
\label{Sionludi}
\end{figure}

\section{Cryogenics}

We use a home-made upside-down dilution refrigerator 
called "sionludi" with the cold plate placed upwards.\cite{AirLiq} 
Figure \ref{Sionludi} shows a schematic of the whole setup.
The cryostat is isolated from the building vibrations by spring 
isolators with a $2 \, Hz$ resonant frequency. The connection to the pumps 
of the dilution circuit is made of flexible pipes and tubes passing 
through heavy concrete blocks. The base 
temperature with the measurement wiring installed is about $60 \, mK$ in a 
cylindrical volume of diameter $13\,cm$ and height $13\,cm$.

The main advantages of our cryostat are its natural rigidity, the 
handy horizontal access and a very rapid cooling/heating cycling. 
The whole cryostat is enclosed in a single vacuum which is sealed with a 
room-temperature o-ring. There is no $N_{2}$ bath and therefore no 
$N_{2}$ boiling vibrations. The $4\, K$ stage is cooled by a 
$^4He$ flow from a pressurized helium vessel. During cooling down to 
$4\, K$, a large flow of $^{3}He-^{4}He$ mixture thermalizes the dilution 
stage. Depending on the $^4 He$ 
flow, the cooling down to $4\, K$ lasts about 6 hours. The $^{3}He-^{4}He$ 
mixture is injected in the cryostat at a pressure between $0.5$ and 
$2 \, bar$. After reaching $4\, K$, a fast cooling down to 
$1.5 \, K$ of the dilution stage is ensured by a Joule-Thomson 
expansion of the mixture through a dedicated circuit. In the dilution 
regime, the limit temperature is reached within 4 hours with the STM 
installed. The cooling power is about $10 \, \mu W$ at $100 \, mK$.

The STM is enclosed in a hermetic calorimeter screwed on the 
mixing chamber. This calorimeter is sealed with an In seal inside a glove box. 
The controlled $N_{2}-He$ dry atmosphere enables us to minimize 
surface contamination. Before cooling, the calorimeter is rinsed 
several times with pure $He$ gas while heating the sample for 
outgassing. The calorimeter is then closed with the help of a stainless 
steel valve with a $He$ pressure of about $10^{-2} \, mbar$. 
At low temperature ($T < 20\, K$), active carbon grains 
pump the residual $He$ atmosphere. 

\begin{figure}
\epsfxsize=8 cm
\epsfbox{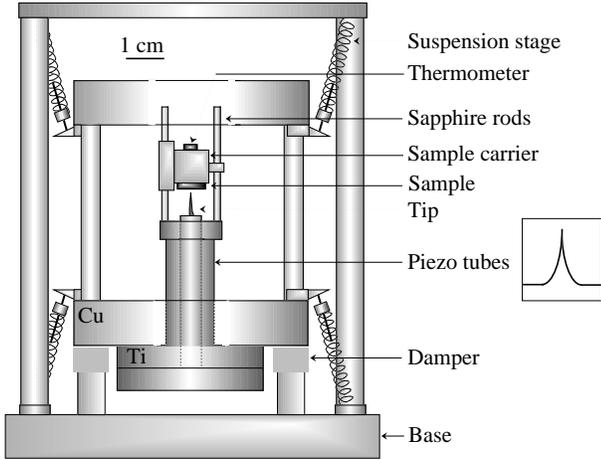}
\caption{Design of the STM head. The two concentric piezoelectric 
tubes are glued with Stycast on Ti parts fixed on the Cu body of the STM. 
The whole STM head is suspended by six springs. The sample carrier 
slides on the two sapphire rods of the outer piezo-electric tube. It 
also includes a special temperature control (see text). 
The coarse approach pulse waveform is schematically shown.}
\label{STM}
\end{figure}

\section{The STM}

\subsection{The STM head}

The STM head is very rigid, $500\, g$ in weight and suspended with 
six springs. The resonant frequency of this isolation stage has been 
adjusted to $10 \, Hz$ in order to avoid any resonance with 
the cryostat isolation stage. The tip is placed near both the inertial 
center of the suspended head and the symmetry center of the spring's fixation 
points. This reduces the displacement of the tip with respect to the 
base. Residual oscillations of the head are damped by sliding friction.

The head is made of two concentric $2.5 \, cm$ long piezo-electric 
PZT-5A tubes. The piezo tubes are glued with Stycast to Ti parts. 
Differential expansion stress during thermal cycles are then minimized 
since Ti and Stycast thermal expansion coefficients are close to 
that of PZT.\cite{Ti} The inner tube holds the tip : it is a XY 
scanner with a $6 \times 6\, \mu m^2$ scan range at low temperature 
for a $\pm \, 220 \, V$ voltage sweep. Its inner electrode is used 
for the withdrawal of the tip during lateral displacements or coarse 
approach steps. The outer piezoelectric tube serves to regulate the sample-tip 
distance regulation and ensures the inertial displacement. It holds the 
two sapphire rods on which the sample inertial carrier slides. 

\begin{figure}
\epsfysize=3.5 cm
\epsfbox{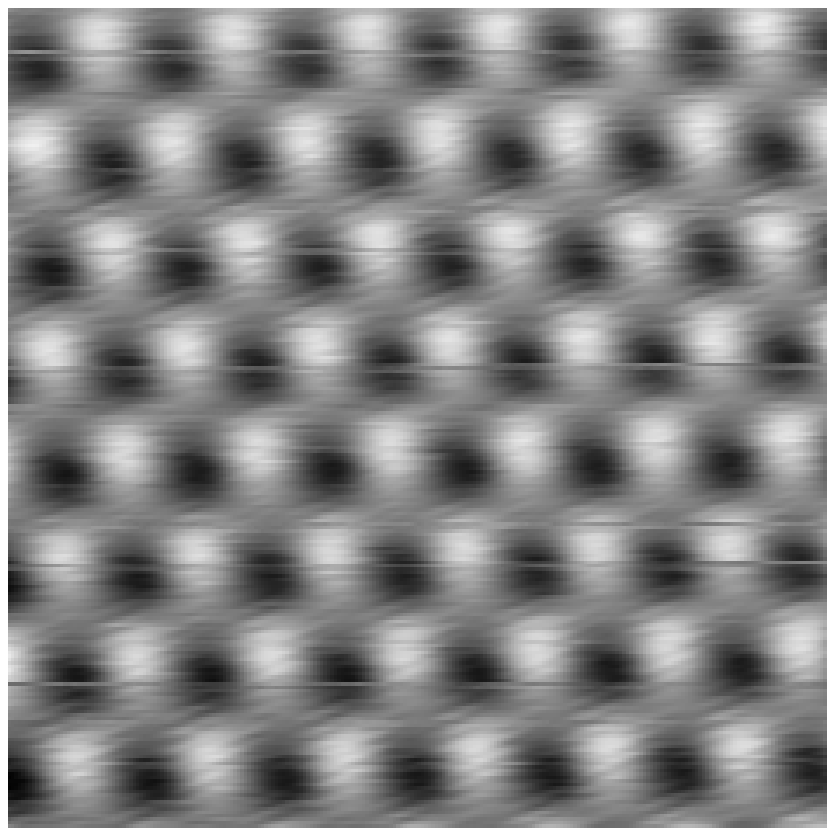}
\epsfysize=3.5 cm
\epsfbox{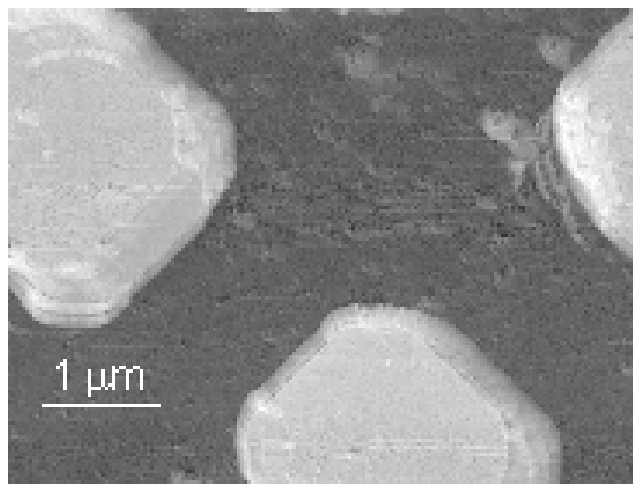}
\caption{Left : $15 \, \times \, 15 \AA^2$ STM image of the atomic 
lattice of $2.4 \, \AA$ periodicity at a HOPG (graphite) surface. Right : 
$5.7 \times 4.3 \, \mu m^2$ STM image of a mesoscopic sample made of a 
square array of $300 \, \AA$-thick Nb square islands with a Au layer 
deposited on top of it. The lattice parameter is $4 \, \mu m$. Both images are 
unfiltered and have been taken at a sample temperature of $60\, mK$ with 
the same piezo-electric tubes.}
\label{Images}
\end{figure}

\subsection{The sample coarse approach}

Our vertical inertial motor is directly inspired from the work of C. Renner 
et al..\cite{Renner} The carrier is made of two brass parts clamped on 
the sapphire rods by a compressed spring. The sliding friction force 
is about $40\, g$ while the carrier weight is $12 \, g$. Each part has three 
well-defined contact points on the rods. The pulse waveform is a double 
parabola, so that the acceleration is constant except at the 
crossover point (see Fig. 2 inset). The duration of one pulse is about 
$240 \, \mu s$. A single pulse makes the carrier move $50$ to $1400 \,nm$ at ambient 
temperature and $10$ to $200\, nm$ at $T < 4\, K$. Both upwards and 
downwards displacement are possible.

We found that the step length remain reliable as long as the sapphire 
tubes remain free of adsorbed water. During cooling, the carrier 
is therefore kept at a temperature about $40\, K$ larger than the cryostat 
temperature. This is ensured by a strain gauge used as a heater 
and a Pt thermometer, both glued on the carrier. This procedure 
ensures a low cryosorption on the sapphire rods and on the sample 
surface. The heating is stopped at a cryostat temperature of $60 \, K$. 

An automatic coarse approach is performed during cooling from $T=60 \, K$ to $4 
\, K$. As the sample is placed about 1 mm above the tip at 
ambient temperature, about 4000 steps are necessary to reach the tunnel 
contact. One step consists in withdrawing the tip with the inner tube, 
sending a pulse on the outer tube and slowly approaching the tip within 5 seconds. 
When a tunnel current is detected, the approach is stopped and 
the tip is withdrawn to a rest 
position. Usually no thermal drift is observed below $20\, K$. 

\subsection{Thermalization} 

The experimental challenge here is 
to ensure both an efficient thermalization and a good mechanical 
decoupling to vibrations for the microscope. 
Thermalization below $1\, K$ is ensured through copper 
braids connecting the base, the suspended head and the Ti parts glued 
on top of the two tubes. The sample 
carrier and the tip are also thermalized through their attached 
Cu measurement wires. A carbon resistor glued on the sample carrier is used 
for the measurement of the sample temperature below $1\, K$.

We have characterized the amount of heating due to both coarse 
approach and scanning. Starting from the STM base temperature 
($60\, mK$), the sample temperature after a single coarse approach 
step reaches instantly about $100\, mK$. It afterwards drops down 
to $60\, mK$ in a few minutes. A continuous displacement with about 
ten steps by second heats the sample up to about $1\, K$ and the mixing 
chamber to $350\, mK$. This corresponds to an injected power of 
about $250 \, \mu W$ and a dissipated energy of about $25 \, \mu J$ per 
step. Image scanning also heats the system but can be minimized 
significantly at low scanning speed. For example, large field 
imaging ($6 \times 6 \, \mu m^2$) on a large corrugation of the order 
of $300\, \AA$ may heat the sample to about $100\, mK$ when the scanning 
speed is about $500\, nm/s$.
 	 
\subsection{The STM electronics}

Our electronics are home-made. Each incoming wires is filtered 
by radio-frequency filters at the cryostat entrance. Switchable very low frequency 
filters with a $0.5\, Hz$ cut-off are placed at the high-voltage 
amplifiers outputs. They are switched on during spectroscopies in 
order to increase stability of the tip-sample distance. The tip bias 
undergoes a $\frac{1}{10}$ reduction at $60 \, mK$ so that the effect 
of extrinsic noise is reduced. The tunnel current 
preamplifier features a $12\, kHz$ bandwidth. It works at room temperature 
but the $100\, M\Omega$ feedback resistance of the input op-amp 
is thermalized at $4.2\, K$. This reduces the contribution of its thermal 
noise to the total current noise. The measured total noise is 
$4.5 fA/\sqrt{Hz}$ below $300\, Hz$ including the current noise from the 
op-amp of about $1.5 fA/\sqrt{Hz}$. Above $300\, Hz$, the noise increases 
linearly with the frequency because the op-amp voltage noise 
biases the capacitance of the coaxial wire. 

\begin{figure}
\epsfxsize=8 cm
\epsfbox{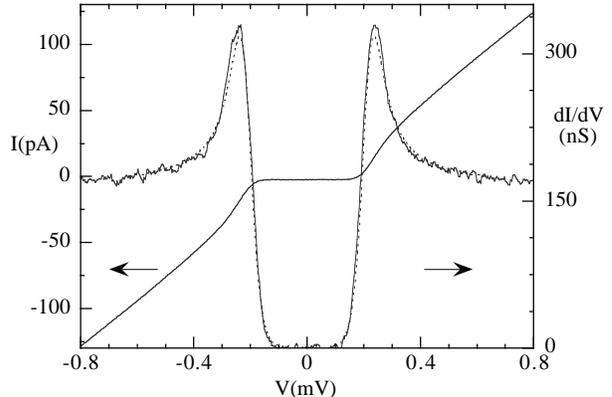}
\caption{I-V characteristics and its numerical derivative (= differential 
conductance) measured by STM on a superconducting layer of Al at 
$60\, mK$. The I-V characteristic is an average over 20 curves. 
The dashed line is a fit using a BCS density of states convoluted with a thermal 
Fermi distribution. The fit parameters are the values of the 
superconducting gap $\Delta_{Al} = 210\,\mu eV$ and an effective temperature 
$T_{eff}= 210 \, mK$.}
\label{SpectroSTM}
\end{figure}

\section{First results}

\subsection{Images}

Mesoscopic samples prepared by standard lithography often exhibit 
large corrugation of several hundreds of $\AA$. With such a large relief, we 
need sharp tips with a large aspect ratio. We use electrochemically-etched Pt/Ir 
tips because of their stiffness and low natural oxidation. Tips are 
prepared with the technique described by Lindahl et 
al.,\cite{Tip} with the difference that we avoid bending the tip but 
instead immerse it in the liquid 
with a large angle. With this method, we obtain routinely a tip radius 
of about $30\, nm$. 

Figure \ref{Images} shows two images taken at the temperature of $60\, mK$. 
Let us point out that the ratio of the image sizes is 
larger than 1000. We achieved atomic resolution on HOPG graphite at both room 
temperature and very low temperature ($60\, mK$). On mesoscopic samples, 
imaging on the scale of a few microns is performed very 
slowly in order to avoid tip crash and heating of the sample. A 
typical $128 \times 128$ pixel image of $5 \times 5 \mu m^2$ is 
acquired in about half an hour. These images show that we are able to 
keep a clean sample surface, i.e., without noticeable contamination, 
during cooling. Heating the sample during cooling was found to be 
necessary in order to obtain such a surface quality.

\begin{figure}
\epsfxsize=8 cm
\epsfbox{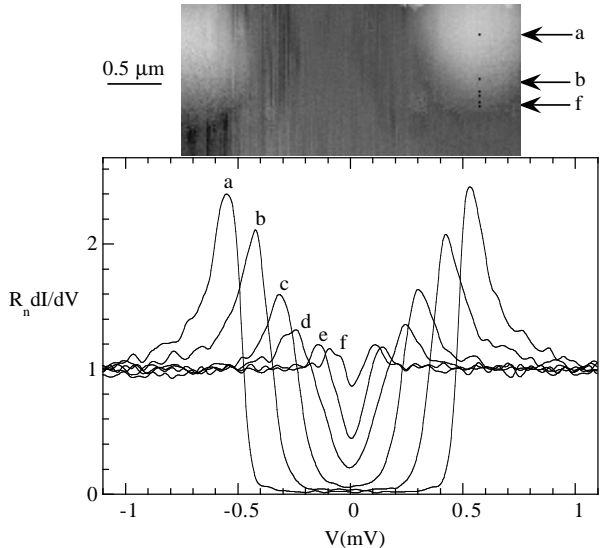}
\caption{Top : $5 \times 2.2 \, \mu m^2$ STM image at $70 \, mK$ of sample 
made of Nb islands covered by an uniform $200\, \AA$ Au layer. The 
locations where the tunnel spectra were measured are indicated by a 
black pixel. Bottom : Measured differential conductance at 
various locations on the sample. At the center of the Nb island, the 
Nb is $440 \, \AA$ thick and the gap is the largest. The tunnel 
conductance was about $300 \, nS$.}
\label{Bulles}
\end{figure}

\subsection{Density of states measurements}

In order to measure a local density of states, 
we acquire I-V characteristics of a fixed sample-tip tunnel junction. 
The tip feedback is first set to a large response time constant, the 
high-voltage low-pass filters are switched on and the feedback loop 
is opened. The bias voltage is then swept while the tunnel current 
is measured. This sweep lasts up to $20\, s$ without modification 
of the junction. We usually average over 10 to 40 I-V curves in 
order to reduce the noise. The experimental I-V characteristic is 
then numerically differentiated in order to obtain the differential 
conductance $dI/dV(V)$. The differential conductance coincides with 
the local density of states in the zero-temperature limit.

As a test experiment, we measured the density of states at the surface 
of plain superconducting layers. Figure \ref{SpectroSTM} shows the 
experimental I-V characteristic and the deduced differential 
conductance $dI/dV$ at the surface of an Al layer. These data are representative of the quality of our spectroscopy 
measurements, although in this case the tip is presumably dug into the surface 
Al oxide. From our measurements, we can estimate a maximum vibration 
amplitude of about $10^{-2} \AA$. The differential conductance 
$dI/dV$ clearly shows a superconducting gap. It is zero in the 
vicinity of the Fermi level ($eV = 0$). We described this data using a 
BCS density of states convoluted by a thermal distribution function, 
without broadening by a finite quasiparticle 
lifetime.\cite{Dynes} The value of the Al gap $\Delta_{Al} = 210\,\mu eV$ 
derived from the fit is in agreement with the expected value. The 
effective temperature derived from the fit is $T_{eff}=210\, mK$. As the sample 
temperature was in every case about $60\, mK$, the origin of this 
effective temperature increase is unclear at present. Nevertheless, our data show that we 
can identify the measured differential conductance with the local 
density of states if the desired resolution remains larger 
than $2k_{B}T_{eff}=36 \mu eV$.

Figure \ref{Bulles} shows a preliminary experiment performed on a 
Normal metal-Superconductor junction. We fabricated the sample by 
successive in-situ evaporation of Nb and Au in a UHV chamber. Nb was 
evaporated first through a Si mask put in contact with the Si 
substrate. The mask was a $5 \, \mu m$ thick Si membrane patterned 
by e-beam lithography. The mask was then mechanically removed and a uniform 
Au layer was deposited, without breaking vacuum. This ensures a high 
transparency of the Nb-Au interface. The 
Nb islands are about $1.8 \, \mu m$ wide and their center to center 
distance is $4.3 \, \mu m$. Only a few Nb islands are accessible within 
the STM scan range. As shown in the STM image, the Nb islands 
are rather rounded. This is due to the thickness of the mask and the 
imperfection of the contact between the mask and the substrate.

We measured the differential conductance at various 
locations while moving the tip away from the center of a Nb island. In order to 
reduce the hysteresis of the scanner below $10 \, nm$, we had to use a 
scanning speed below $10 \, nm/s$. In Fig. \ref{Bulles}, we clearly see 
the spatial dependence of the density of states. The spectrum with the 
largest gap was taken near the center of the Nb island : it can be 
well described by a BCS density of states. 
The magnitude of the energy gap is reduced compared to the Nb value 
because of the influence of the Au layer. Away from the Nb island, 
the density of states has a triangular shape. It is 
the signature of the proximity effect. The characteristic 
decay length of the proximity effect appears to be about $120 \, nm$. 
A detailed description of our results by the quasiclassical 
theory \cite{Belzig} is under way.

This work greatly beneficiated from the 
help of A. Benoit, H. Rodenas and K. Hasselbach in the design, 
fabrication and optimization of the "sionludi" cryostat. We thank the 
CRTBT technical staff for their various contributions to the design and 
construction of the STM setup. We thank W. Belzig, A. Bezryadin, I. 
Maggio-Aprile, C. Renner, D. Rodichev, H. Takayanagi and Y. de 
Wilde for discussions. We acknowledge discussions within the "Dynamics 
of superconducting nanocircuits" TMR network and support from DRET 
and R\'egion Rh\^one-Alpes.

\end{document}